\documentclass[%
 reprint,
 amsmath,amssymb,
 aps,
]{revtex4-1}

\usepackage{graphicx}
\graphicspath{ {images/} }
\usepackage{dcolumn}
\usepackage{bm}

\usepackage[usenames,dvipsnames]{color}

\begin{document}


\title{Ferroelectricity in [111]-oriented epitaxially strained SrTiO$_3$ from first principles}

\author{Sebastian E. Reyes-Lillo,$^{(1,2,3)}$ Karin M. Rabe,$^{(4)}$ and Jeffrey B. Neaton$^{(2,3,5)}$}
\affiliation{
(1) Departamento de Ciencias F\'isicas, Universidad Andres Bello, Santiago 837-0136, Chile  \\
(2) Molecular Foundry, Lawrence Berkeley National Laboratory, Berkeley, California 94720, USA \\
(3) Department of Physics, University of California, Berkeley, California 94720, USA \\
(4) Department of Physics and Astronomy, Rutgers University, Piscataway, New Jersey 08854, USA \\
(5) Kavli Energy NanoSciences Institute at Berkeley, Berkeley, California 94720, USA
}\date{\today}

\begin{abstract}
We use first principles density functional theory calculations to investigate the effect of biaxial strain in the low-temperature structural and ferroelectric properties of [111]-oriented SrTiO$_3$. We find that [111] biaxial strain, achievable by coherent epitaxial growth along the [111] direction, induces structural distortions in SrTiO$_3$ that are not present in either bulk or [001]-oriented SrTiO$_3$. Under [111] biaxial strain, SrTiO$_3$ displays ferroelectricity at tensile strain, and paraelectricity at compressive strain. We compute the phonon spectrum and macroscopic polarization of SrTiO$_3$ as a function of [111] biaxial strain, and relate our results to the predictions of the free energy phenomenological model of Pertsev, Tagantsev and Setter [Phys. Rev. B 61, 825 (2000); Phys. Rev. B 65, 219901 (2002)].
\end{abstract}

\pacs{Valid PACS appear here}
\maketitle

Advances in molecular beam epitaxy have prompted tremendous interest in the effect of coherent epitaxial strain on transition metal oxides~\cite{Dawber2005, Rabe2007}. Biaxial strain of [001]-oriented SrTiO$_3$ ultrathin films, associated with epitaxial growth on cubic perovskite substrates, has been extensively studied and predicted to provide routes to significantly tune its structural~\cite{Pertsev1998}, electronic~\cite{Berger2011}, and dielectric~\cite{Antons2005} properties. However, the effects of biaxial strain associated with perovskite films grown along the [111] direction, on substrates with hexagonal symmetry, have received less attention. 

The structural effect of [111] biaxial strain on ABO$_3$ perovskites has been previously investigated by focusing on a dominant polar or antiferrodistortive instability. First principles calculations predict ferroelectricity for [111]-oriented PbTiO$_3$~\cite{Oja2008} and BaTiO$_3$~\cite{Raeliarijaona2017, Oja2008}, and the existence of Goldstone-like modes in the structural energy landscape of [111]-oriented LaAlO$_3$~\cite{Moreau2017} and SrMnO$_3$~\cite{Marthinsen2018}. Recently, employing an automated computational workflow based on first principles methods, Angsten and co-workers studied the effect of [001], [011] and [111] biaxial strain on the polar instability of a large set of perovskite oxides~\cite{Angsten2017}. In addition, (001) and (111) surfaces of SrTiO$_3$ and other perovskites were recently studied from first principles~\cite{Eglitis2008, Eglitis2015, Eglitis2015a}. Here, motivated by recent experimental and theoretical studies on ferroelectric~\cite{Xu2018a, Li2007}, electronic~\cite{Doennig2013}, magnetic~\cite{Hallsteinsen2016, Piamonteze2015, Tybell2015, Gibert2012}, transport~\cite{Dix2012} and topological~\cite{Marthinsen2018, Patzner2016, Weng2015, Xiao2011} phenomena on [111]-oriented perovskites and superlattice interfaces, we examine the case of perovskites with simultaneous polar and antiferrodistortive structural instabilities, such as SrTiO$_3$.

SrTiO$_3$ is a prototypical ABO$_3$ perovskite that, under certain conditions, display important technological properties such as photocatalysis~\cite{Wrighton1976} and superconductivity~\cite{Schooley1964, Kozuka2009}. In bulk form, SrTiO$_3$ has a cubic \textsl{Pm$\bar{3}$m} structure at temperatures above 105~K and a tetragonal \textsl{I4/mcm} structure at low temperatures below 105~K~\cite{Shirane1969} characterized by oxygen octahedron rotations (a$^0$a$^0$c$^-$ in Glazer notation). The frequency of the lowest-frequency polar mode is seen to decrease with temperature with an extrapolated Curie temperature of about $\sim$30K~\cite{Muller1979}, but this incipient ferroelectricity~\cite{Uwe1976} is suppressed by a combination of the oxygen octahedron rotations in the tetragonal phase ~\cite{Zhong1995, Yamanaka2000, Sai2000} and quantum fluctuations at low temperature ~\cite{Muller1979, Zhong1996}. 

First-principles calculations of the phonon dispersion of cubic SrTiO$_3$ show multiple lattice instabilities~\cite{Desquesnes2005}. An unstable antiferrodistortive zone-boundary $R_5^-$ mode, corresponding to concerted rotation of the oxygen octahedra, gives rise to the low-temperature tetragonal structure, here denoted as \textsl{I4/mcm}($R_5^-$[001]). An unstable zone-center polar $\Gamma_4^-$ mode can be associated with the incipient ferroelectricity. Additional phonon instabilities, including finite-wavevector octahedral rotations along the $R-M$ direction of the Brillouin zone, can lower the energy of the cubic structure in principle, but are not observed in bulk phases.

The coupling of lattice instabilities to epitaxial strain can alter the energy landscape and stabilize novel phases, including a rich variety of ferroelectric phases with or without nonpolar antiferrodistortive distortions~\cite{Pertsev2000, Pertsev2002}. Phenomenological models~\cite{Thomas1968, Pertsev2000, Pertsev2002} and first principles calculations~\cite{Dieguez2005, Hashimoto2005, Lin2006} predict the emergence of ferroelectricity in [001]-oriented perovskite films under epitaxial strain. The emergence of ferroelectricity is readily explained by strain-polarization coupling, in which ferroic off-centering of the Ti atoms in each unit cell arises along elongated lattice vector directions. In the free energy expansion of Pertsev and co-workers~\cite{Pertsev2000}, this is manifested in the term $-(S_{1}P_1^2+S_{2}P_2^2+S_{3}P_3^2)$ ; $\{P_i\}_{i=1,2,3}$ and $\{S_i\}_{i=1-6}$ denote the polarization and strain (Voigt notation) components.  At compressive (tensile) [001] biaxial strain: $S_1=S_2<0$ and $S_3>0$ ($S_1=S_2>0$ and $S_3<0$), and the system lowers its energy by adopting the $P_1=P_2=0$ and $P_3 \neq 0$ ($P_1=P_2 \neq 0$ and $P_3 = 0$) configuration.

First-principles calculations show increased instability of the polar mode with elongation along a particular direction, and predict the stabilization of the ferroelectric phases \textsl{P4mm}($\Gamma_4^-$[001]) and \textsl{Amm2}($\Gamma_4^-$[110]) at compressive and tensile [001] biaxial strain, respectively~\cite{Ni2011a}. Indeed, thin film SrTiO$_3$ is observed to be ferroelectric above critical compressive and tensile [001] biaxial strains, and to posses a strain-tunable Curie temperature~\cite{Haeni2004}.

Allowing oxygen octahedral rotations increases the energy gain of polar phases and lower their symmetry. Under [001] biaxial strain, the free energy term $-(S_{1}Q_1^2+S_{2}Q_2^2+S_{3}Q_3^2)$, where $\{Q_i\}_{i=1,2,3}$ denotes out-of-phase  octahedral rotations along cubic lattice directions, favors out-of-plane (in-plane) octahedral rotations at compressive (tensile) strain and the system lowers its energy by adopting the $Q_1=Q_2=0$ and $Q_3 \neq 0$ ($Q_1=Q_2 \neq 0$ and $Q_3 = 0$) configuration. Consequently, first principles calculations predict the stabilization of the polar phases \textsl{I4cm}($\Gamma_4^-$[001]+$R_5^-$[001]) and \textsl{Ima2}($\Gamma_4^-$[110]+$R_5^-$[110]) at compressive and tensile strain, respectively.

The phenomenological approach can also be applied to the case of biaxial strain along the [111] direction induced by epitaxy. In this case, the strain-polarization coupling term takes the form $-S(P_1^2+P_2^2+P_3^2)$ due to the symmetry constraint $S=S_1=S_2=S_3$ in the reference structure. In contrast to the [001] biaxial strain case where ferroelectricity is favored at both tensile and compressive strain, the later coupling predicts ferroelectric ($|\vec{P}| \neq 0$) and antiferrodistortive ($|\vec{P}| = 0$) distortions at tensile ($S>0$) and compressive ($S<0$) [111] biaxial strains, respectively. In addition, the off-diagonal shear strain components $S_4$, $S_5$ and $S_6$, which typically cancel in the case of [001] growth direction, assume non zero values and can modify the relative energy stability of polar and antiferrodistortive distortions, as shown below.

\begin{figure}[htp]
\includegraphics[scale=0.35]{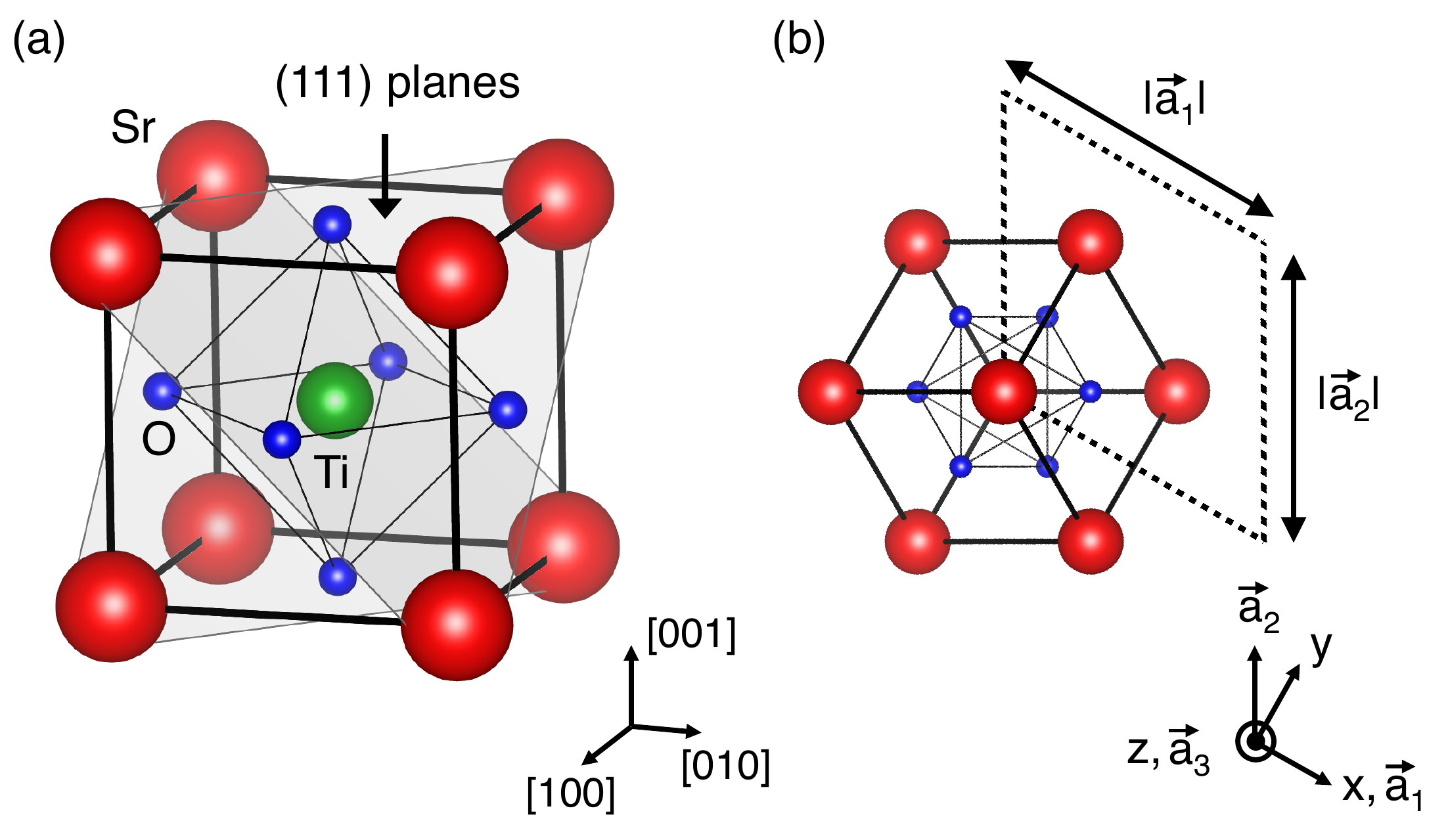}
\caption{\label{fig:fig1} (a) Cubic unit cell displaying a (111) plane and cubic lattice vectors ([100], [010], [001]). (b) Relationship between hexagonal ($\vec{a}_1$, $\vec{a}_2$, $\vec{a}_3$) and Cartesian ($x$, $y$, $z$) lattice vectors. Sr, Ti and O atoms are denoted by red, green and blue spheres, respectively.}
\end{figure}

In this paper, we use symmetry arguments and first principles calculations to investigate the effect of [111] biaxial strain on the structural and ferroelectric properties of SrTiO$_3$. We find that, similar to the case of [001] biaxial strain, [111]-oriented SrTiO$_3$ displays several low-symmetry phases with polar ferroelectric and nonpolar antiferrodistortive distortions that are not present in bulk form. However, unlike the [001] case, where SrTiO$_3$ becomes ferroelectric at both tensile and compressive strain, [111] biaxially strained SrTiO$_3$ becomes ferroelectric and paraelectric at tensile and compressive strain, respectively. We list all symmetry-allowed ferroelectric phases under [111] biaxial strain and investigate their behavior under strain. Our first principles results are supported by the predictions of a free-energy phenomenological model and can be generalized to other perovskite oxides.

\begin{figure}[htp]
\includegraphics[width=\columnwidth]{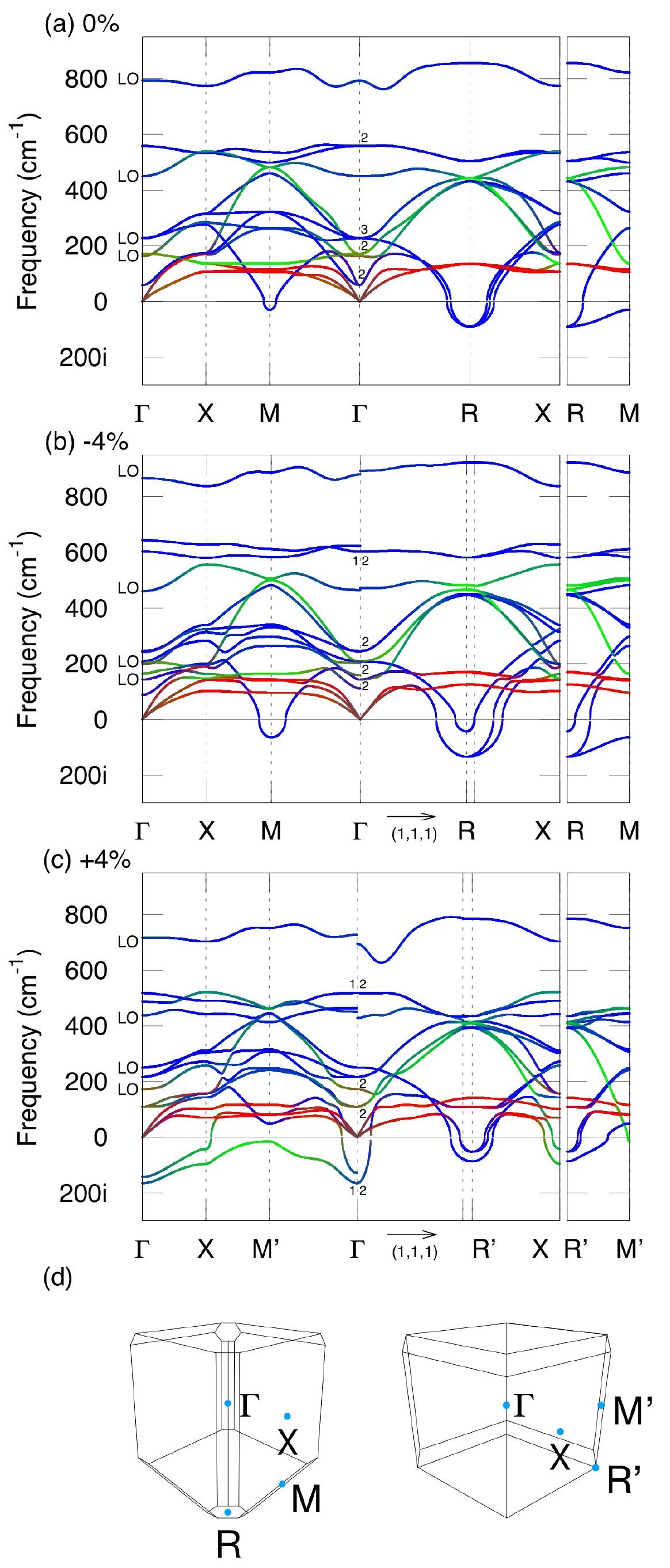}
\caption{\label{fig:fig2} Phonon dispersion for SrTiO$_3$ under (a) $0$\%, (b) $-4$\% and (c) $+4$\% [111] biaxial strain computed with DFT-LDA. Colors denote the atomic displacement magnitude for Sr (red), Ti (green), O (blue) at a given phonon frequency. We label longitudinal optical (LO) modes and phonon degeneracies. For easier comparison, we use the symmetry point labels $\Gamma=(0,0,0)$, $X=(1/2,0,0)$, $M=(1/2,1/2,0)$, $M'=(1/2,-1/2,0)$, $R=(1/2,1/2,1/2)$ and $R'=(1/2,-1/2,1/2)$. (d) Brillouin zone for the reference structure at $-4$\% (left) and $+4$\% (right) strain~\cite{Hinuma2017}. Nonzero [111] biaxial strain lowers the symmetry of the Brillouin zone, and leads to an additional symmetry line around the cubic $R$ point.} 
\end{figure}

\begin{table}[b]
\caption{DFT calculations are performed in hexagonal supercells. For each supercell we report the number of formula units (\# f.u.), lattice vector lengths, and $\vec{k}$-point grids used in total-energy (polarization) calculations.}
\begin{ruledtabular}
 \begin{tabular}{lccc} 
 \# f.u. &  $|\vec{a}_1|=|\vec{a}_2|$ & $|\vec{a}_3|$ &  $\vec{k}$-point grid   \\ 
\hline
3                 &    $\sqrt{2}a_0$    & $\sqrt{3}a_0$       &      $5 \times 5 \times 5$ ($9 \times 9 \times 9$)   \\ 
6                 &    $\sqrt{2}a_0$    & $2\sqrt{3}a_0$     &      $5 \times 5 \times 3$  ($9 \times 9 \times 7$) \\ 
24               &    $2\sqrt{2}a_0$  & $2\sqrt{3}a_0$    &      $3 \times 3 \times 3$ ($7 \times 7 \times 7$)   \\ 
 \end{tabular}
 \end{ruledtabular}
\label{table:dft}
\end{table}

Density functional theory (DFT) calculations are performed, within the local density approximation (LDA)~\cite{Perdew1981}, using the \emph{Vienna Ab-initio Simulation Package}~(VASP)~\cite{Kresse1996a, Kresse1996}. As shown in Fig.~\ref{fig:fig1}, epitaxial growth on the (111) surfaces of perovskite substrates is simulated through strained bulk calculations~\cite{Pertsev1998, Dieguez2005} in hexagonal cells. Misfit strain is measured with respect to the computed cubic lattice parameter $a_0=3.863$\AA. Structural relaxations are performed, keeping the matching plane lattice vectors $\vec{a}_1$ and $\vec{a}_2$ fixed ($\sphericalangle_{\vec{a}_1, \vec{a}_2}=120^{\circ}$) while the out-of-plane lattice vector $\vec{a}_3$ and the internal atomic positions are optimized until forces are smaller than $0.1$~meV/\AA. 

Table~\ref{table:dft} shows the $\vec{k}$-point grid and lattice vector lengths for the hexagonal cells considered in this paper. Our \textsc{vasp} calculations use $\Gamma$ centered $\vec{k}$-point grids and a plane wave energy cut-off of $500$~eV. Projector augmented wave pseudopotentials~\cite{Kresse1999c} explicitly include 10 valence electrons for Sr($4s^24p^65s^2$), 12 for Ti ($3s^23p^63d^24s^2$) and 6 for O ($2s^22p^4$). Phonon dispersions are computed in the primitive cell using finite differences, $4 \times 4 \times 4$ supercell structures with $2 \times 2 \times 2$ $\Gamma$ centered $\vec{k}$-point grids, and are interpolated with~\textsc{phonopy}~\cite{Togo2015}. Berry phase polarization calculations are performed within the modern theory of polarization, as implemented in~\textsc{vasp}~\cite{King-Smith1993}. Our DFT convergence parameters provide good agreement with previous results~\cite{Birol2011, Himmetoglu2014a}. 

\begin{table}[t]
\caption{For each experimental bulk phases of SrTiO$_3$ we report the space group, energy (meV/f.u.) with respect to cubic \textsl{Pm$\bar{3}$m}, and lattice parameters (\AA). Experimental lattice parameters (in parenthesis) are taken from Refs.~\cite{Schmidbauer2012, Jauch1999}}
\begin{ruledtabular}
 \begin{tabular}{lcc} 
 Space  group &  Energy  (meV/f.u.) &   Latt.  parameters (\AA) \\ 
\hline
\textsl{Pm$\bar{3}$m}     &          0           & $a_0=3.863 (3.905)$       \\ 
\textsl{I4/mcm}                  &       $-$11          & $a=b=5.442 (5.507)$                     \\
                                            &                       &  $c=7.753 (7.796)$                    \\
 \end{tabular}
 \end{ruledtabular}
\label{table:bulk}
\end{table}

Table~\ref{table:bulk} shows the energy and lattice parameters for bulk phases of SrTiO$_3$. Computed lattice parameters underestimate experimental lattice constants by $\sim$1\%, as is typical for the LDA. The tetragonal \textsl{I4/mcm} structure corresponds to the ground state, in agreement with experiments.    
The [111] biaxial strain constraint lowers the symmetry of the cubic perovskite \textsl{Pm$\bar{3}$m} reference structure to \textsl{R$\bar{3}$m}. We start by considering the effects of this symmetry lowering on the phonon dispersion of the reference \textsl{R$\bar{3}$m} structure, shown for  $0$\%, $-4$\%, and $+4$\% [111] biaxial strain in Fig.~\ref{fig:fig2}. Our calculations for $0$\% strain are in line with previous results for cubic SrTiO$_3$ using the LDA~\cite{Birol2011, Himmetoglu2014a}. The polar transverse optical (TO) mode $\Gamma_4^-$ is stable but soft ($62$ $cm^{-1}$), and the antiferrodistortive $M_2^+$ and $R_5^-$ modes are unstable ($28i$ $cm^{-1}$ and $91i$ $cm^{-1}$, respectively) (see Fig.~\ref{fig:fig2}(a)).

At $-$4\% (compressive) strain, the phonon dispersion is quite similar to that at $0$\%: The lowest frequency polar $\Gamma_4^-$ mode is slightly stiffened  ($142$ $cm^{-1}$) and the antiferrodistortive instabilities at $M$ and $R$ are slightly enhanced ($65i$ $cm^{-1}$ and $135i$ $cm^{-1}$, respectively) (see Fig.~\ref{fig:fig2}(b)). The symmetry lowering produces a visible splitting of the unstable modes at $R$ and of the lowest frequency polar $\Gamma_4^-$ mode, as well as the slight dependence of the LO mode frequencies and splittings at $\Gamma$ on the direction of $\vec q$ in the limit $q \rightarrow 0$.

In contrast, $+4$\% (tensile) strain has a dramatic effect on the phonon dispersion. In addition to a suppression of the antiferrodistortive instabilities at $M$ and $R$ ($24i$ $cm^{-1}$ and $86i$ $cm^{-1}$, respectively) (see Fig.~\ref{fig:fig2}(c)), we find a strong polar instability at $\Gamma$ ($170i$ $cm^{-1}$), comparable to the instability in [001] biaxial strain that generates the ferroelectric phases. This lowest phonon branch is unstable all the way from $\Gamma$ to $X$, a characteristic feature also of the phonon dispersion of cubic BaTiO$_3$~\cite{Ghosez1999}.

\begin{table*}
\caption{For each bulk phase, we report the oxygen octahedral tilting (in Glazer notation), symmetry mode decomposition, space group symmetry under [111] biaxial strain, number of formula units (f,.u.) and space group symmetry obtained upon freezing in the polar $\Gamma_4^-$ mode along different cubic directions. Symmetry modes are given with respect to cubic \textsl{Pm$\bar{3}$m} with atomic positions given by Sr (1a), Ti (1b), and O (3c). The polar structure \textsl{P}1   $(\pm p_1 \pm p_2 \pm p_3)$ is allowed in every entry. }
\begin{ruledtabular}
 \begin{tabular}{lccccccccc} 
   Bulk           & Glazer     & Symm. &  Strained & f.u. &                                Polar phases                                            \\ 
  phase       & notation  &   modes     &   phase     & &                                                                                                                       \\ 
\hline
 \textsl{Pm$\bar{3}$m}   & a$^0$a$^0$a$^0$ &     & \textsl{R$\bar{3}$m} & 3  &  \textsl{R3m} $\pm(ppp)$;  \textsl{C2}   $\pm(p\bar{p}0)$, $\pm(p0\bar{p})$, $\pm(0p\bar{p})$;  \textsl{Cm}   $\pm(p_1p_1\pm p_2)$, $\pm(p_1\pm p_2p_1)$, $\pm(\pm p_2p_1p_1)$             \\ 
 \textsl{I4/mcm}               & a$^0$a$^0$c$^-$  &   R$_5^-$  & \textsl{C2/c}  & 6   &   \textsl{C2}   $\pm(p\bar{p}0)$, $\pm(p0\bar{p})$, $\pm(0p\bar{p})$;  \textsl{Cc}   $\pm(p_1p_1\pm p_2)$, $\pm(p_1\pm p_2p_1)$, $\pm(\pm p_2p_1p_1)$              \\  
 \textsl{Imma}                  & a$^0$b$^-$b$^-$   &   R$_5^-$   & \textsl{C2/c}  & 6   &   \textsl{C2}   $\pm(p\bar{p}0)$, $\pm(p0\bar{p})$, $\pm(0p\bar{p})$;  \textsl{Cc}   $\pm(p_1p_1\pm p_2)$, $\pm(p_1\pm p_2p_1)$, $\pm(\pm p_2p_1p_1)$            \\ 
 \textsl{R$\bar{3}$c}      & a$^-$a$^-$a$^-$    &   R$_5^-$  & \textsl{R$\bar{3}$c}  & 6   &  \textsl{R3c} $\pm(ppp)$;  \textsl{C2}  $\pm(p\bar{p}0)$, $\pm(p0\bar{p})$, $\pm(0p\bar{p})$;  \textsl{Cc}   $\pm(p_1p_1\pm p_2)$, $\pm(p_1\pm p_2p_1)$, $\pm(\pm p_2p_1p_1)$            \\  
 \textsl{P4/mbm}             & a$^0$a$^0$c$^+$ &   M$_2^+$                       & \textsl{P2$_1$/c}  & 6   &    \textsl{P2$_1$}  $\pm(p\bar{p}0)$, $\pm(p0\bar{p})$, $\pm(0p\bar{p})$;   \textsl{Pc}   $\pm(p_1p_1\pm p_2)$, $\pm(p_1\pm p_2p_1)$, $\pm(\pm p_2p_1p_1)$   \\ 
 \textsl{I4/mmm}             & a$^0$b$^+$b$^+$  &   M$_2^+$                      & \textsl{C2/m}      & 6       &    \textsl{C2}   $\pm(p\bar{p}0)$, $\pm(p0\bar{p})$, $\pm(0p\bar{p})$;  \textsl{Cm}   $\pm(p_1p_1\pm p_2)$, $\pm(p_1\pm p_2p_1)$, $\pm(\pm p_2p_1p_1)$   \\  
 \textsl{Im$\bar{3}$}      & a$^+$a$^+$a$^+$  &   M$_2^+$                       & \textsl{R$\bar{3}$} & 6   &            \textsl{R3} $\pm(ppp)$                  \\ 
 \textsl{C2/m}                  & a$^0$b$^-$c$^-$   &    R$_5^-$                        & \textsl{P$\bar{1}$} & 6   &                            \\  
  \textsl{Pnma}               & a$^+$b$^-$b$^-$    &    R$_5^-$, M$_2^+$    & \textsl{P2$_1$/c} & 6    &  \textsl{P2$_1$}  $\pm(p\bar{p}0)$, $\pm(p0\bar{p})$, $\pm(0p\bar{p})$;   \textsl{Pc}   $\pm(p_1p_1\pm p_2)$, $\pm(p_1\pm p_2p_1)$, $\pm(\pm p_2p_1p_1)$  \\ 
 \textsl{C2/c}                  & a$^-$b$^-$b$^-$     &    R$_5^-$                   & \textsl{C2/c}  & 6      &   \textsl{C2}   $\pm(p\bar{p}0)$, $\pm(p0\bar{p})$, $\pm(0p\bar{p})$;  \textsl{Cc}   $\pm(p_1p_1\pm p_2)$, $\pm(p_1\pm p_2p_1)$, $\pm(\pm p_2p_1p_1)$                                         \\  
  \textsl{Cmcm}                    & a$^0$b$^+$c$^-$        &              R$_5^-$, M$_2^+$      & \textsl{P$\bar{1}$}   & 6        &                          \\ 
 \textsl{P4$_2$/nmc}          & a$^+$a$^+$c$^-$        &              R$_5^-$, M$_2^+$       & \textsl{C2/c}  & 6  &      \textsl{C2}   $\pm(p\bar{p}0)$, $\pm(p0\bar{p})$, $\pm(0p\bar{p})$;  \textsl{Cc}   $\pm(p_1p_1\pm p_2)$, $\pm(p_1\pm p_2p_1)$, $\pm(\pm p_2p_1p_1)$                     \\  
  \textsl{Immm}                     & a$^+$b$^+$c$^+$       &              M$_2^+$                        & \textsl{P$\bar{1}$}  & 6        &                            \\ 
 \textsl{P2$_1$/m}              & a$^+$b$^-$c$^-$         &              R$_5^-$, M$_2^+$       & \textsl{P$\bar{1}$}    & 6      &                             \\  
 \textsl{P$\bar{1}$}             & a$^-$b$^-$c$^-$          &              R$_5^-$                          & \textsl{P$\bar{1}$}  & 6         &                              \\  
  \end{tabular}
 \end{ruledtabular}
\label{table:hexagonal}
\end{table*}

In the following, we investigate the low-symmetry structures compatible with the reference \textsl{R$\bar{3}$m} structure, obtained by freezing in the unstable cubic modes ($\Gamma_4^-$, $R_5^-$, and $M_2^+$). Table~\ref{table:hexagonal} shows bulk, strained, and polar phases of SrTiO$_3$. For each cubic rotational pattern, we report the nature of the oxygen octahedral tilting (in Glazer notation), symmetry mode decomposition, space group symmetry under [111] biaxial strain, number of formula units, and polar phases obtained by freezing in the polar $\Gamma_4^-$ mode along different cubic directions (see Fig.~\ref{fig:fig1}). For each polar structure, we include the free parameters of the polarization vector in the cubic setting. We list polar phases that are invariant under at least one symmetry operation and group polarization directions into symmetry equivalent structures (variants).

Under [111] biaxial strain, \textsl{Pm$\bar{3}$m} is reduced to \textsl{R$\bar{3}$m}, and the commonly known polar phases of BaTiO$_3$ and PbTiO$_3$ --\textsl{R3m}, \textsl{Amm2} and \textsl{P4mm}-- evolve smoothly to \textsl{R3m}, \textsl{C2} or \textsl{Cm}~\cite{Oja2008}. The [001]-oriented ferroelectric phases with eight (\textsl{R3m}), twelve (\textsl{Amm2}), and six (\textsl{P4mm}) variants rearrange into [111]-oriented ferroelectric phases with two (\textsl{R3m}), six (\textsl{C2}) and three sets of six (\textsl{Cm}) variants~\cite{Tagantsev2001}. Similarly, the \textsl{I4/mcm} structure lowers its symmetry to monoclinic \textsl{C2/c} ($a^-b^-b^-$ in Glazer notation) and leads to the ferroelectric phases \textsl{C2} and \textsl{Cc}, combining polar and antiferrodistortive distortions.

Table~\ref{table:hexagonal} includes hypothetical bulk phases. Similar to \textsl{I4/mcm}, the \textsl{Imma} and \textsl{P4$_2$/nmc} structures lower their symmetry to monoclinic \textsl{C2/c}. The \textsl{C2/c} structure, characterized by having the same rotational pattern along two different cubic directions, leads to the ferroelectric phases \textsl{C2} and \textsl{Cc}.  Freezing in the polar $\Gamma_4^-$ mode in the \textsl{R$\bar{3}$c} phase leads to the ferroelectric phases \textsl{R3c}, \textsl{C2} (denoted as \textsl{C2}$^{\ast}$ in the following to distinguish it from the \textsl{C2} structure without octahedral tilting) and \textsl{Cc}. The \textsl{Pnma} structure, the ground state of several oxide perovskites~\cite{Eklund2009}, lowers its symmetry to \textsl{P2$_1$/c} and leads to the polar structures \textsl{P2$_1$} and \textsl{Pc}. The \textsl{P4/mbm}, \textsl{I4/mmm}, and \textsl{Im$\bar{3}$} structures lower their symmetry to \textsl{P2$_1$/c}, \textsl{C2/m}, and \textsl{R$\bar{3}$}, respectively. Finally, several phases, \textsl{C2/m}, \textsl{Cmcm}, \textsl{Immm}, \textsl{P2$_1$/m}, and \textsl{P$\bar{1}$}, lead to the low-symmetry \textsl{P$\bar{1}$} structure.

\begin{figure}[b]
\includegraphics[width=\columnwidth]{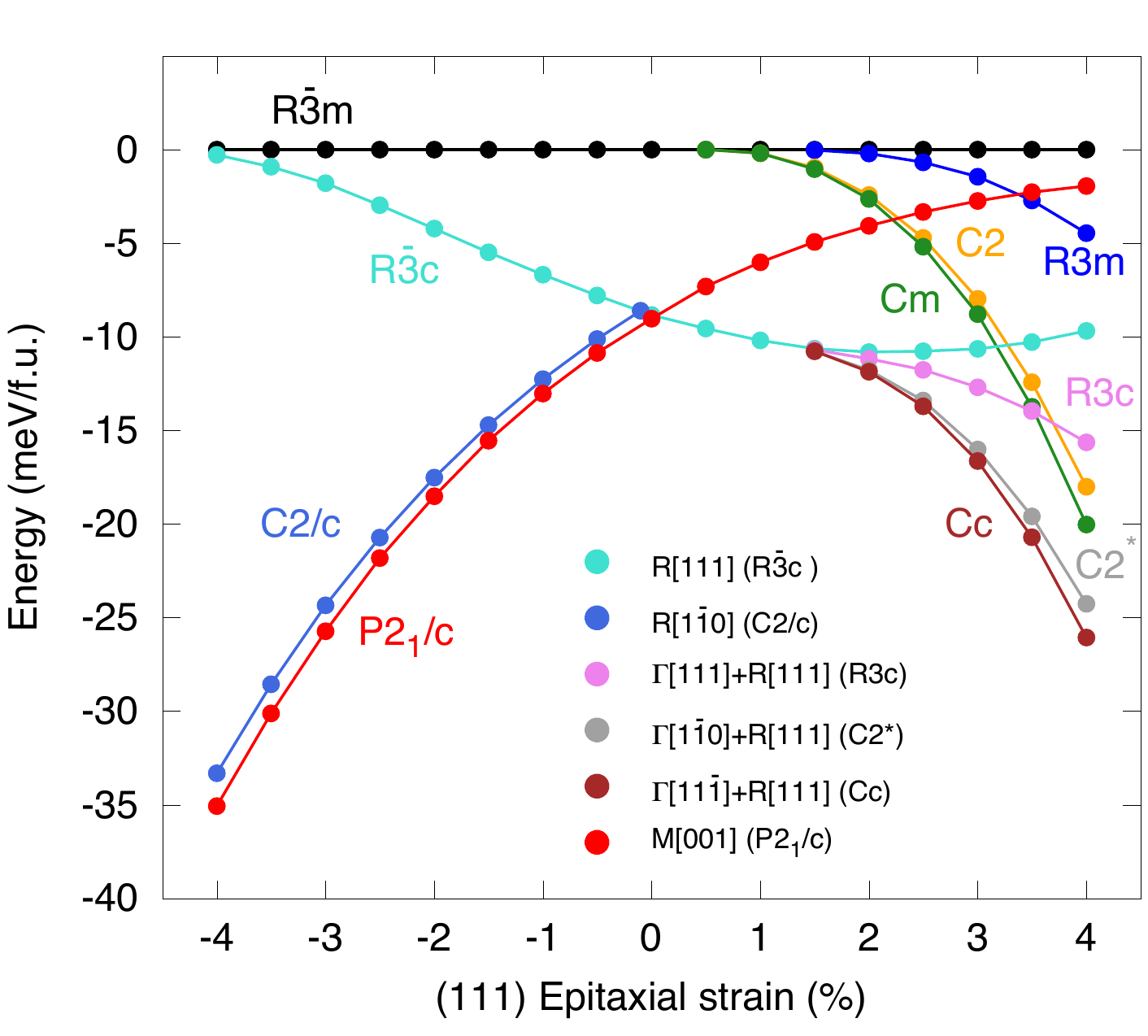}
\caption{\label{fig:fig3} Energy (meV/f.u.) versus [111] biaxial strain (\%) diagram for SrTiO$_3$. For each strain value, relative energy stability is computed with respect to the reference \textsl{R$\bar{3}$m} structure. Negative (positive) values represent compressive (tensile) [111] biaxial strain. The legend shows the symmetry modes for the low-symmetry structures. \textsl{C2$^*$} denotes the structure obtained from \textsl{C2} by freezing in the $R_5^-$[111] mode.}.
\end{figure}

Next, we explore the relative energy stability of the structures in Table~\ref{table:hexagonal} under [111] biaxial strain. Figure~\ref{fig:fig3} shows the energy gain with respect to \textsl{R$\bar{3}$m}. The strain diagram displays several low-symmetry low-energy structures. At zero-temperature, nonpolar \textsl{P2$_1$/c} is favored under compressive strain. Under tensile strain, two phases --nonpolar \textsl{R$\bar{3}$c} at small strain, and then polar \textsl{Cc} at higher strain-- are predicted. At tensile strain, several polar phases are nearby in energy and display a small energy difference with \textsl{R$\bar{3}$m}, suggesting that is quite likely to observe a ferroelectric phase at higher temperatures and large enough strain. We organize our following analysis by looking at the free energy coupling terms involving polar, antiferrodistortive and strain degrees of freedom. We refer to Pertsev and co-workers~\cite{Pertsev2000} for the sign and relative magnitude of the coefficients in front of the free-energy terms.

Similar to the [001] biaxial strain case, antiferrodistortive oxygen octahedral rotations are stabilized in the entire strain range due to the strong $R_5^-$ instability. At compressive strain, the paraelectric phases \textsl{C2/c} and \textsl{P2$_1$/c} dominate at low energy. At $-4$\% strain, the rotational angles along the cubic directions are given by $11^\circ$, $6^\circ$, and $6^\circ$ for \textsl{C2/c} (a$^-$b$^-$b$^-$ in Glazer notation) and $0^\circ$, $11^\circ$ and $11^\circ$ for \textsl{P2$_1$/c} (a$^+$b$^-$b$^-$ in Glazer notation). The unstable M$_2^+$ cubic mode, not included in the phenomenological model, breaks additional symmetries and favors \textsl{P2$_1$/c} over \textsl{C2/c} by a small energy difference. At tensile strain, the paraelectric phase \textsl{R$\bar{3}$c} (a$^-$a$^-$a$^-$) corresponds to the ground state energy structure between $0$\% and $+1.2$\% strain and is later suppressed above $\sim +1.5$\% strain.

The splitting of the $R_5^-$ mode suppresses the antiferrodistortive \textsl{R$\bar{3}$c} phase (a$^-$a$^-$a$^-$) at compressive strain  (see Fig.~\ref{fig:fig3}). For rotational patters away from the [111] axis, the coefficient in front of the free energy term $+S(Q_1^2+Q_2^2+Q_3^2)$ ($S=S_1=S_2=S_3$) is positive and predicts enhancement and suppression of octahedral tiltings at compressive ($S<0$) and tensile ($S>0$) strains, respectively. Accordingly, the rotational angles decrease with strain to $8$ ($0$\%) and $5$ ($4$\%) for \textsl{P2$_1$/c}, and to $10^\circ$, $4^\circ$, and $4^\circ$ ($0$\%) for \textsl{C2/c}. For \textsl{R$\bar{3}$c}, the rotational angles increase with strain, and are given by $2^\circ$, $6^\circ$ and $7^\circ$ at $-4$, $0$, and $4$\% strains, respectively.

Interestingly, the splitting of the three fold $R_5^-$ cubic mode induces a first-order transition between \textsl{P2$_1$/c} and \textsl{R$\bar{3}$c}, as well as a discontinuity in the energy profile of \textsl{C2/c}. The splitting of the $R_5^-$ mode and the discontinuity of the rotational patterns originates from the free energy term $+(S_4 Q_2 Q_3 + S_5 Q_1 Q_3 + S_6 Q_1 Q_2)$. Under strain, this term takes the form $+S_v(Q_2 Q_3 + Q_1 Q_3 + Q_1 Q_2)$ ($S_v=S_4=S_5=S_6$) and approaches $-S_v Q^2$ and $+S_v 3Q^2$ for \textsl{C2/c}($R_5^-$[1$\bar{1}$0]) and \textsl{R$\bar{3}$c}($R_5^-$[111]), respectively, favoring \textsl{C2/c} and \textsl{R$\bar{3}$c} at compressive ($S_v>0$) and tensile ($S_v<0$) strain, respectively. Above $0$\% strain, \textsl{C2/c} converges to the higher-symmetry structure \textsl{R$\bar{3}$c} and displays a discontinuity.

In sharp contrast to the [001] biaxial strain case, all ferroelectric phases are stabilized at tensile [111] biaxial strain, in agreement with the results of Fig.~\ref{fig:fig2}. The free-energy term $-S(P_1^2+P_2^2+P_3^2)$ ($S=S_1=S_2=S_3$) has a negative sign and predicts suppression and enhancement of ferroelectricity at compressive ($S<0$) and tensile ($S>0$) strains, respectively. The ferroelectric phases \textsl{Cm}, \textsl{C2}, and \textsl{R3m} are stabilized at $+0.5$\%, $+0.5$\%, and $+1.3$\% strains, respectively, due to the stabilization of the $\Gamma_4^-$ phonon mode. 

The relative strain stabilization of \textsl{Cm}, \textsl{C2}, and \textsl{R3m} at tensile strain and the splitting of the unstable TO mode at $\Gamma$ (see Fig.~\ref{fig:fig2}(c)) originates by the free energy term $-(S_4 P_2 P_3 + S_5 P_1 P_3 + S_6 P_1 P_2)$. Upon stabilization of the polar $\Gamma_4^-$ mode at tensile strain, the coupling term takes the form $-S_v(P_2 P_3 + P_1 P_3 + P_1 P_2)$ ($S_v=S_4=S_5=S_6$), and approaches $+S_v P^2$ for \textsl{Cm}($\Gamma_4^-$[11$\bar{1}$]) and \textsl{C2}($\Gamma_4^-$[1$\bar{1}$0]), and $-S_v 3P^2$ for \textsl{R3m}($\Gamma_4^-$[111]). The later effectively favors \textsl{Cm} and \textsl{C2} over \textsl{R3m} by a small energy difference. As shown in Fig.~\ref{fig:fig3}, in the absence of oxygen octahedral rotations, the lowest energy structure corresponds to \textsl{Cm}~\cite{Angsten2017}.

Polar and antiferrodistortive coupling terms further increase the energy gain of the ferroelectric phases. The \textsl{R3c}, \textsl{C2}$^{\ast}$, and \textsl{Cc} structures, combining polar and antiferrodistortive distortions ($\Gamma_4^-$ and $R_5^-$ instabilities), are stabilized almost simultaneously at $+1.2$\%, $+1.1$\% and $+1.2$\% strain, respectively. The strain-independent free energy term $-(P_1 P_2 Q_1 Q_2 + P_1 P_3 Q_1 Q_3 + P_2 P_3 Q_2 Q_3)$ shifts the critical strain stabilization of the ferroelectric phases, and above $+1.2$\% strain, favors the ferroelectric \textsl{Cc} phase as the ground state of the system. Notably, the low-symmetry polar phases \textsl{P2$_1$}, \textsl{Pc}, \textsl{R3}, and \textsl{P1} are not stabilized in SrTiO$_3$ and converge to the higher-symmetry \textsl{Cc} phase. 

\begin{figure}[h]
\includegraphics[width=\columnwidth]{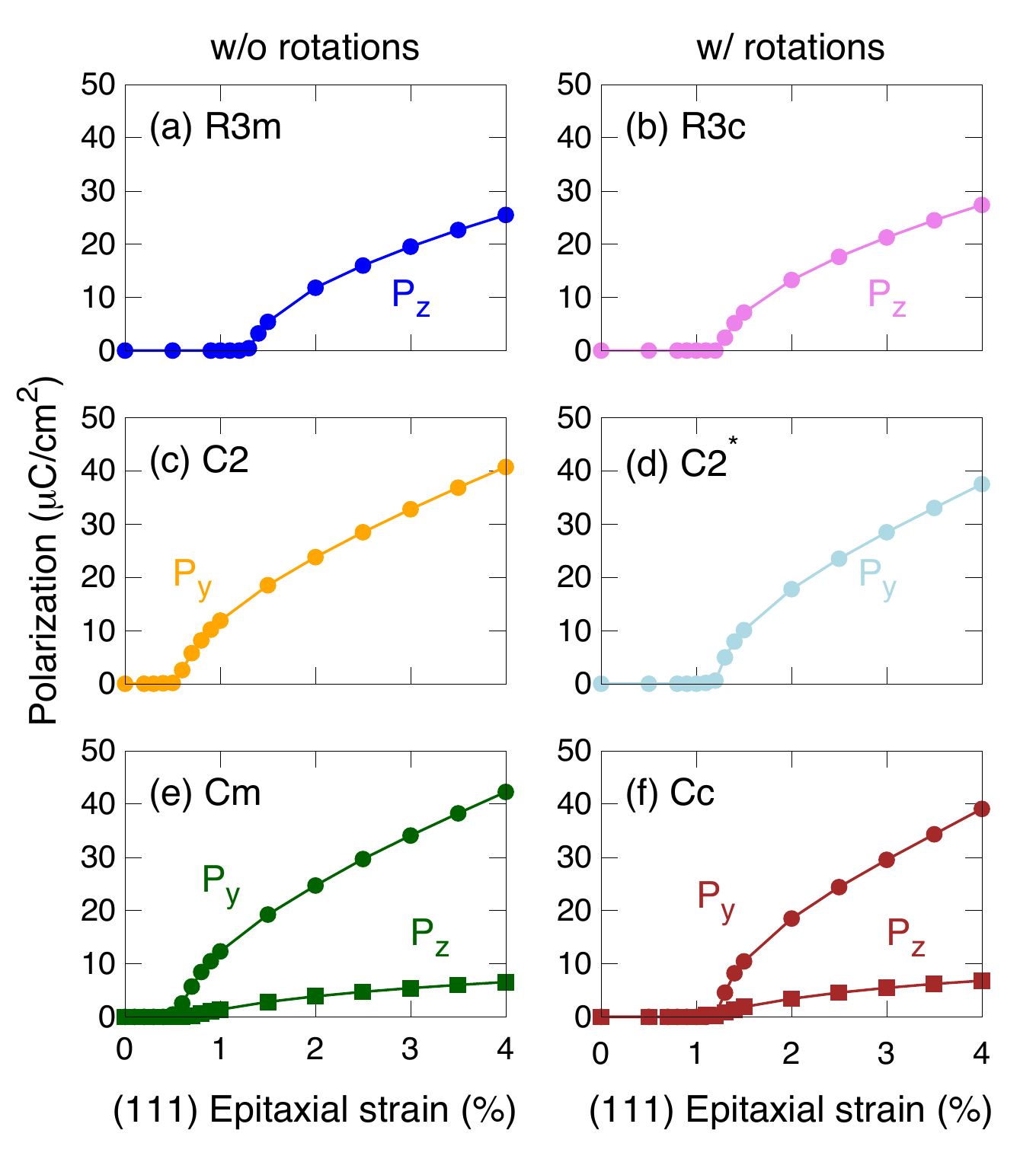}
\caption{\label{fig:fig4} Polarization ($P$) versus [111] biaxial strain along the Cartesian coordinates (${x,y,z}$). The relation between Cartesian and cubic directions is given in Fig.~\ref{fig:fig1}. Polar structures are separated among those without: (a) \textsl{R3m}, (b) \textsl{C2}, (c) \textsl{Cm}, and those with: (d) \textsl{R3c}, (e) \textsl{C2}$^{\ast}$, (f) \textsl{Cc}, oxygen octahedral rotations.}
\end{figure}

Finally, Fig.~\ref{fig:fig4} shows the polarization as a function of [111] biaxial strain for the ferroelectric phases shown in Fig.~\ref{fig:fig3}. Polarization components are shown along Cartesian and cubic directions in Fig.~\ref{fig:fig4} and Table~\ref{table:hexagonal}, respectively, and their relation is described in Fig.~\ref{fig:fig1}. Similar to the [001] case, polarization becomes nonzero above a critical strain, and its magnitude increases with tensile strain. The maximum polarization value at $+4$\% [111] biaxial strain corresponds to 42$~\mu C/cm^2$ for \textsl{Cm} ($7$ and $42~\mu C/cm^2$ in-plane and out-of-plane, respectively), smaller than the maximum polarization value computed for an equivalent amount of [001] biaxial strain ($53$ and $50~\mu C/cm^2$ for \textsl{P4mm} and \textsl{Amm2} at compressive and tensile [001] biaxial strains, respectively). 

Ferroelectric phases display in-plane and/or out-of-plane polarization components at tensile [111] biaxial strain. Notably, the out-of-plane polarization component of \textsl{Cm} can be induced by an in-plane electric field due to a symmetry allowed third-order term in the free energy expansion~\cite{Tagantsev2001}, suggesting a coupling mechanism between in-plane and out-of-plane polarization components. Our results are expected to be in qualitative agreement with experiments, and in particular, to be independent of the choice of DFT functional. In this regard, we find that the generalized gradient approximation~\cite{Perdew1996} leads to qualitatively similar results than the LDA, and predicts \textsl{Cc} as the ground state structure at tensile strain.

More generally, the symmetry conditions imposed on the material by [111] biaxial strain are summarized by the strain conditions $S=S_1=S_2=S_3$ and $S_v=S_4=S_5=S_6$ in the reference coordinates, with $S_v \ll S$. The latter shows that biaxial strain applied perpendicular to the [111] growth direction resembles the application of hydrostatic pressure. Unlike [001] biaxial strain, which favors Ti off centering along out- and in-plane lattice vectors, tensile [111] biaxial strain induces a uniform elongation of the unit cell and favors Ti off centering simultaneously along the out- and in-plane directions. Therefore, tensile [111] biaxial strain softens the TO mode and favors the emergence of ferroelectricity.

Our first-principles results can be easily generalized to other ABO$_3$ perovskite oxides with polar and antiferrodistortive instabilities. In particular, the symmetry analysis of Table~\ref{table:hexagonal} is valid for any bulk or layered perovskite heterostructure. However, the relative stability of the symmetry-allowed nonpolar and polar phases will depend on the stoichiometric details of the material or, equivalently, the coefficients in front of the free-energy terms in the phenomenological model. 

In summary, DFT calculations show that SrTiO$_3$ display ferroelectricity at tensile [111] biaxial strain and paraelectricity at compressive [111] biaxial strain. At tensile strain, ferroelectricity emerge with the softening of the lowest TO polar mode and polarization increases with tensile strain. At compressive strain, the absence of ferroelectricity can be explained by a homogeneous suppression of the polar mode.  

\section{\label{sec:level1} Acknowledgments}

We thank D. Schlom, B. Montserrat, R. F. Berger, and T. Birol for valuable discussions. This work is supported by the U.S. Department of Energy, Office of Science, Office of Basic Energy Sciences, Materials Sciences and Engineering Division under Contract No. DE-AC02-05CH11231: Materials Project program No. KC23MP. This research used resources of the National Energy Research Scientific Computing Center, a DOE Office of Science User Facility supported by the Office of Science of the U.S. Department of Energy under Contract No. DE-AC02-05CH11231. Additional calculations were performed at the Molecular Foundry,  supported by the Office of Science, Office of Basic Energy Sciences, under Contract No. DE-AC02-05CH11231. The work of K. M. R. was supported by Office of Naval Research N00014-17-1-2770. The work of S. E. R. was partly supported by Fondo Nacional de Desarrollo Cient\'ifico y Tecnol\'ogico (FONDECYT, Chile) through program Iniciaci\'on en Investigaci\'on 2018 under grant No. 11180590.


%
%

\end{document}